\documentclass[11pt]{article}
\usepackage{blois,epsfig}
\usepackage{graphicx}

\def\be{\begin{equation}}
\def\ee{\end{equation}}
\def\bea{\begin{eqnarray}}
\def\eea{\end{eqnarray}}

\begin{document}
\vspace*{4cm}
\title{Recent Tau Decay Results at B Factories $\sim$Lepton Flavor Violating Tau Decays$\sim$}

\author{ K.Hayasaka }

\address{Nagoya university, Furo-cho,\\
Chikusa-ku, Nagoya, Japan}

\maketitle\abstracts{
We report recent results for lepton flavor violating (LFV) $\tau$
decays obtained at Belle and BaBar with the world-largest data samples.
}

\section{Introduction}
An observation of the lepton flavor violation (LFV) 
is a clear signature of New Physics (NP) since
the LFV in charged leptons has a negligibly 
small probability in the Standard Model (SM)
even taking into account neutrino oscillations. 
For example, 
the branching fraction for the $\tau$ decay into $\mu$ and $\gamma$,
should be proportional to squared mass  difference between 
neutrinos and the weak boson mass squared and is
calculated to be less than $10^{-54\mbox{--} -40}$.~\cite{Pham:1998fq}
Since
the $\tau$ lepton can have many possible LFV decay modes
and is expected to be sensitive to NP because
it has the heaviest mass among charged leptons,
we search for
the LFV from the $\tau$ lepton as a NP phenomenon.

The model including supersymmetry (SUSY), 
that is the most popular 
as the one beyond the SM,
can naturally induce LFV at one loop.
In many SUSY models, a $\tau\rightarrow\mu\gamma$
decay is expected to have the largest 
branching fraction among all the $\tau$ LFV decays
while a branching fraction for
a $\tau$ decay into $\mu\eta$ or $\mu\mu\mu$
can become the largest in some cases, 
such as the higgs-mediated case.
By evaluating the branching fractions for various 
$\tau$ LFV decays, we can distinguish what model is 
favored. Here, we especially focus on the three major modes,
$\tau\rightarrow\ell \gamma$, $\ell \ell' \ell''$ and $\ell P^0$
where $\ell=e$ or $\mu$ and $P^0=\pi^0$, $\eta$ or $\eta'$.

The world-highest
$\tau$ data samples has been collected in
two B-factory experiments, by the Belle and BaBar collaborations,
because the cross section for $b\bar{b}$
is very close to that for $\tau^+\tau^-$ on the $\Upsilon(4S)$ mass.
Therefore, the B-factory is a good place to search
for the $\tau$ LFV. Here, recent results with ${\cal O}(10^9)$
$\tau$ samples obtained by
Belle and BaBar are reported.

\section{Analysis Method}

\begin{figure}[h]
\begin{center}
\includegraphics[width=0.7\textwidth]{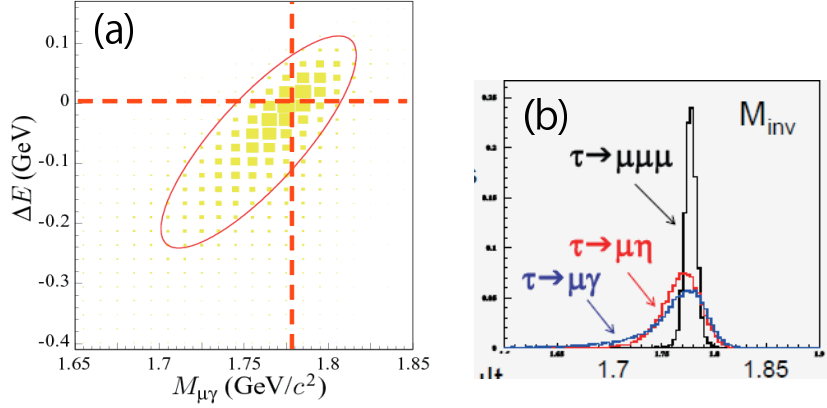}
\end{center}
 \caption{ (a) Distribution for the $\tau\rightarrow\mu\gamma$ signal
 events. (b) Comparison for the resolutions of the reconstructed $\tau$
 mass for $\tau\rightarrow\mu\gamma$ (blue), $\tau\rightarrow\mu\eta$
 (red) and $\tau\rightarrow\mu\mu\mu$ (black). }
\label{fig:signalMCmg}
\end{figure}

In the $\tau$ LFV analysis, in order to evaluate the number of
 signal events, two independent variables are defined,
that are
signal-reconstructed mass
and energy in the center-of-mass (CM) frame
from energies and momenta for the signal $\tau$ daughters. 
In the $\tau\rightarrow\mu\gamma$ case,
they are defined as
\begin{eqnarray}
 M_{\mu\gamma}=\sqrt{E_{\mu\gamma}^2-P_{\mu\gamma}},\\
 \Delta E=E_{\mu\gamma}^{\rm CM}-E_{\rm beam}^{\rm CM},
\end{eqnarray}
where $E_{\mu\gamma}$ ($P_{\mu\gamma}$) is a sum of the energies 
(a magnitude of a vector sum of the momenta) for $\mu$ and $\gamma$,
the superscript $\rm CM$ indicates that the variable is defined in the
CM frame and the $E_{\rm beam}^{\rm CM}$ 
means the initial beam energy in the CM frame.
Principally, $M_{\mu\gamma}$ and $\Delta {E}$ should be
$m_{\tau}$ ($\sim 1.78$ GeV/$c^2$) and 0 (GeV), respectively,
for signal events while $M_{\mu\gamma}$ and $\Delta {E}$ will
smoothly vary without any special structure in the background (BG)
events. (BaBar often takes an energy-constrained mass
$(M_{\rm EC})$ rather than the usually reconstructed mass.)
The signal MC distribution of $\tau\rightarrow\mu\gamma$
 on the $M_{\mu\gamma}$ --
$\Delta{E}$ plane is shown in Fig.~\ref{fig:signalMCmg}.
Due to the resolution, the signal events are distributed
 around $M_{\mu\gamma}\sim m_{\tau}$ 
and $\Delta E\sim0$ (GeV).
Taking into account the resolution, we set the elliptic signal region.
(See Fig.~\ref{fig:signalMCmg}.)
Finally,
we evaluate the number of signal events in the signal region.
Generally, when some gammas are included, the resolution become
worse. Therefore, $\tau\rightarrow\ell\ell'\ell''$ has
a good resolution while $\tau\rightarrow\ell\gamma$ has
a bad resolution. This means that good (bad) $S/N$ is
expected in the  $\tau\rightarrow\ell\ell'\ell''$ 
($\tau\rightarrow\ell\gamma$) analysis because the number of
the BG events can be roughly estimated to be proportional
to the size of the signal region.

To avoid any bias for our analyses, we perform the blind analysis:
Before fixing the selection criteria and the evaluation 
for the systematic uncertainties, we cover the data 
events in the signal region.

\section{\boldmath $\tau\rightarrow\ell\gamma$}
In 2008, Belle has updated their result for $\tau\rightarrow\ell\gamma$
with $4.9\times10^8$ $\tau^+\tau^-$-pairs.
\cite{Belle:mg} The main BG events come from
$\tau\rightarrow\ell\nu\nu$ + extra $\gamma$ events
and the radiative di-muon or bhabha events.
The number of the signal events is evaluated by
a generalized extended unbinned maximum-likelihood to the
$M_{\mu\gamma}$--$\Delta E$ distribution.
As a result, the upper limit on the branching
fraction for $\tau\rightarrow\mu\gamma$ $(e\gamma)$
is set to be $4.5\times10^{-8}$ ($1.2\times10^{-7}$) 
at 90\% confidence level (CL).
Recently, BaBar has published the new result
with $4.8\times10^8$ $\tau^+\tau^-$-pairs data sample
on $\Upsilon(4S)$ as well as $\Upsilon(2S)$ and 
$\Upsilon(3S)$ masses. New requirements to especially reject
radiative di-muon or bhabha events are introduced.
In the $2\sigma$ signal region,
The signal detection efficiency for $\tau\rightarrow\mu\gamma$
($e\gamma$) is 6.1\% (3.9\%) while the expected number of
the BG events is $6.1\pm0.5$ ($3.9\pm0.3$) events.
Consequently, 2 (0) events are found 
in the $\tau\rightarrow\mu\gamma$ ($e\gamma$) analysis
and the 90\% upper limit on the branching fraction
is evaluated to be 4.4  $(3.3)\times10^{-8}$.
Here, the 90\% CL upper limits are calculated by a counting method.
\begin{figure}[h]
\begin{center}
\includegraphics[width=0.6\textwidth]{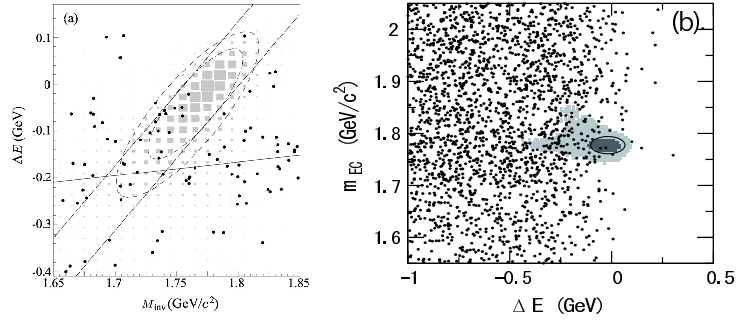}
\end{center}
\caption{Resulting distributions for $\tau\rightarrow\mu \gamma$
of Belle~(a) and BaBar~(b). The
black dots and the shaded boxes show 
the data and signal MC distributions, respectively,
and the (most inside) elliptic region
is the signal region.}
\end{figure}

\section{\boldmath $\tau\rightarrow\ell\ell'\ell''$}

In the last summer, Belle and BaBar have shown
the results for $\tau\rightarrow\ell\ell'\ell''$
with their almost full data samples, of
$7.2\times10^8$ $\tau^+\tau^-$-pairs by Belle~\cite{Belle:3l} and 
$4.3\times10^8$ ones by BaBar~\cite{BaBar:3l}.
BaBar improves the efficiency for the lepton-ID and
the better rejection for the background is obtained.
As a result, no event in the signal region is found
for all modes and the 90\% CL upper limits on the branching 
fractions are set as shown in Table~\ref{table:3l}.
On the other hand, Belle takes very similar
selection criteria to those for the previous analysis.
Finally, for all modes, no event is observed in the signal region 
and the upper limits are evaluated around 2 or 3 times more
sensitively than those obtained previously. 
\begin{table}[h]
\footnotesize
\begin{tabular}[t]{|l|c|c|c||l|c|c|c|}
\hline
 Mode ($\tau^-\rightarrow$) & Eff.(\%) &$N_{BG}^{\rm exp}$ & UL
 ($\times10^{-8}$)& 
 Mode ($\tau^-\rightarrow$) & Eff.(\%) &$N_{BG}^{\rm exp}$ & UL
 ($\times10^{-8}$)\\
\hline
$e^-e^+e^-$ &6.0&$0.21\pm0.15$&2.7&   $e^-\mu^+\mu^-$ &6.1&$0.10\pm0.04$&2.7\\
            &8.6&$0.12\pm0.02$&3.4&   $e^-\mu^+\mu^-$ &6.4&$0.54\pm0.14$&4.6\\
\hline
$e^-e^+\mu^-$ &9.3&$0.04\pm0.04$&1.8& $\mu^-e^+\mu^-$ &10.1&$0.02\pm0.02$&1.7\\
              &8.8&$0.64\pm0.19$&3.7& $\mu^-e^+\mu^-$ &10.2&$0.03\pm0.02$&2.8\\
\hline
$e^-\mu^+e^-$ &11.5&$0.01\pm0.01$&1.5& $\mu^-\mu^+\mu^-$&7.6&$0.13\pm0.06$&2.1\\
              &12.6&$0.34\pm0.12$&2.2& $\mu^-\mu^+\mu^-$ &6.6&$0.44\pm0.17$&4.0\\
\hline
\end{tabular}
\caption{Summary for the efficiency (Eff.),
the expected number of the BG events, ($N_{BG}^{\rm exp}$)
and the upper limit on the branching fraction (UL) for each mode.
The upper and lower lines show the Belle and BaBar results,
respectively. }
\label{table:3l}
\end{table}
\begin{figure}[h]
\begin{center}
\includegraphics[width=0.6\textwidth]{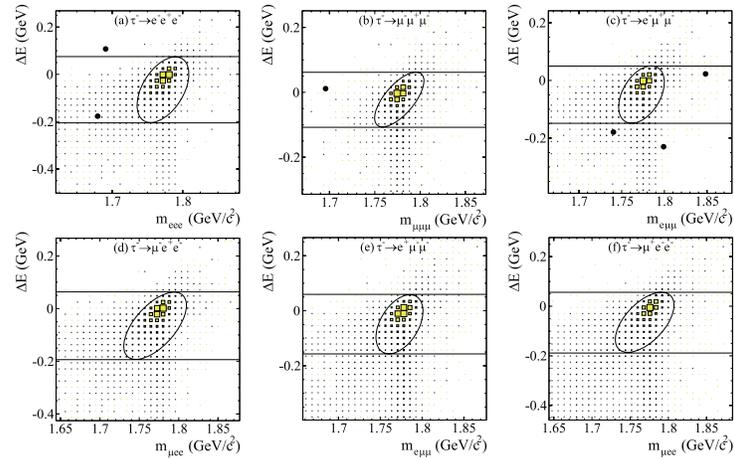}
\end{center}
\caption{Resulting distributions for $\tau\rightarrow\ell\ell'\ell''$
of Belle.
The black dots and the shaded boxes show 
the data and signal MC distributions, respectively,
and the elliptic region is
the signal region.}
\end{figure}
\section{\boldmath $\tau\rightarrow\ell P^0$ $(P^0=\pi^0, \eta, \eta')$}
\begin{figure}[h]
\includegraphics[width=0.9\textwidth]{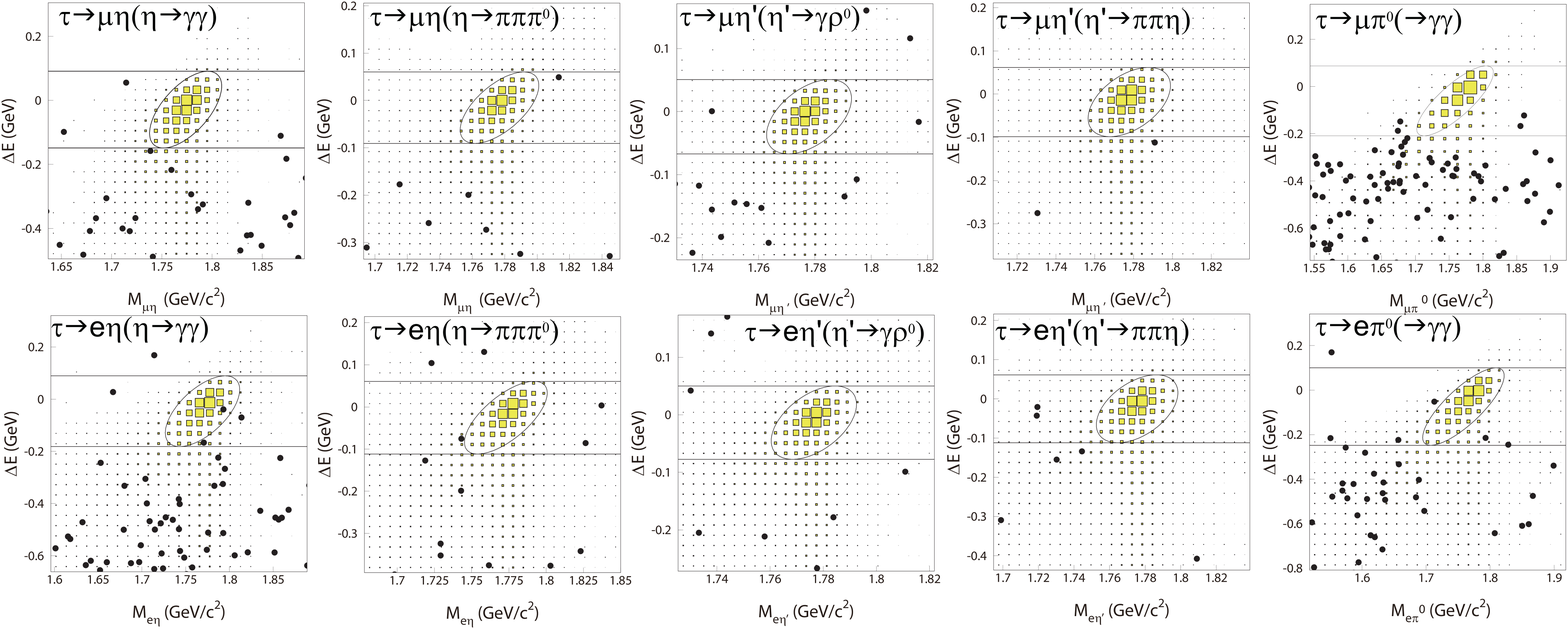}
\vspace*{-2mm}
\caption{Resulting distributions for $\tau\rightarrow\ell P^0$. The
black dots and the yellow boxes show 
the data and signal MC distributions, respectively,
and the elliptic region
is the signal region for each mode. This is preliminary.}
\label{fig:lp0}
\end{figure}

In 2007, Belle and BaBar have published their
results of the search for the $\tau$ decay into a lepton
and a neutral pseudoscalar $(\pi^0, \eta, \eta')$.
They have used around 400fb${}^{-1}$ and
set the range of the upper limits of $(0.8-2.4)\times10^{-7}$
at 90\% CL.~\cite{Belle:leta}~\cite{BaBar:leta} 
This summer, Belle reports the new result
with about 900fb${}^{-1}$.
By studying the backgrounds in detail, 
they obtain about 1.5 times larger efficiency
than previously in average and
the expected number of the BG events in the signal
region is achieved to be less than one for each mode. 
Details are summarized in Table.~\ref{table:lp0}.
Finally, one event is found in 
$\tau\rightarrow e\eta(\rightarrow\gamma\gamma)$
while no event is observed in other modes.
(See Fig.~\ref{fig:lp0})
The evaluated 90\% upper limits on the 
branching fraction are $(2.2-4.4)\times10^{-8}$.

\begin{table}[h]
\footnotesize
\begin{tabular}[t]{|l|c|c|c||l|c|c|c|}
\hline
 Mode ($\tau\rightarrow$) & Eff.(\%) &$N_{BG}^{\rm exp}$ & UL
 ($\times10^{-8}$)& 
 Mode ($\tau\rightarrow$) & Eff.(\%) &$N_{BG}^{\rm exp}$ & UL
 ($\times10^{-8}$)\\
\hline
$\mu\eta(\rightarrow\gamma\gamma)$ &8.2&$0.63\pm0.37$&3.6&
$e\eta(\rightarrow\gamma\gamma)$ &7.0&$0.66\pm0.38$&8.2\\
\hline
$\mu\eta(\rightarrow\pi\pi\pi^0)$ &6.9&$0.23\pm0.23$&8.6&
$e\eta(\rightarrow\pi\pi\pi^0)$ &6.3&$0.69\pm0.40$&8.1\\
\hline\hline
$\mu\eta$(comb.) &&&2.3&
$e\eta$(comb.) &&&4.4\\
\hline\hline
$\mu\eta'(\rightarrow\pi\pi\eta)$ &8.1&$0.00^{+0.16}_{-0.00}$
&10.0&
$e\eta'(\rightarrow\pi\pi\eta)$ &7.3&$0.63\pm0.45$&9.4\\
\hline
$\mu\eta'(\rightarrow\gamma\rho^0)$ &6.2&$0.59\pm0.41$&6.6&
$e\eta'(\rightarrow\gamma\rho^0)$ &7.5&$0.29\pm0.29$&6.8\\
\hline\hline
$\mu\eta'$(comb.) &&&3.8&
$e\eta'$(comb.) &&&3.6\\
\hline\hline
$\mu\pi^0$ &4.2&$0.64\pm0.32$&2.7&
$e\pi^0$ &4.7&$0.89\pm0.40$&2.2\\
\hline
\end{tabular}
\vspace*{-2mm}
\caption{Summary for the efficiency (Eff.),
the expected number of the BG events, ($N_{BG}^{\rm exp}$)
and the upper limit on the branching fraction (UL) for each mode,
where (comb.) means the combined result from subdecay modes.}
\vspace*{-5mm}
\label{table:lp0}
\end{table}

\section{Summary}
We report recent results of the various search for LFV $\tau$ decays
obtained by Belle and BaBar. These results are about 100 times
more sensitive than those by CLEO. This is achieved by 
not only much larger data samples but more
 effective BG rejection after detailed examination of the BG.
At the present, we are updating results using
full data samples accumulated by B-factories.


\begin{thebibliography}{99}
\bibitem{Pham:1998fq}
  X.~Y.~Pham,
  Eur.\ Phys.\ J.\  C {\bf 8}, 513 (1999).
\bibitem{Belle:mg}
  K.~Hayasaka {\it et al.}  (Belle Collaboration),
  Phys.\ Lett.\  B {\bf 666}, 16 (2008).
\bibitem{BaBar:mg}
  B.~Aubert {\it et al.}  (BaBar Collaboration),
  Phys.\ Rev.\ Lett.\  {\bf 104}, 021802 (2010).
\bibitem{Belle:3l}
  K.~Hayasaka {\it et al.} (Belle Collaboration),
  Phys.\ Lett.\  B {\bf 687}, 139 (2010).
\bibitem{BaBar:3l}
  J.~P.~Lees {\it et al.}  (BaBar Collaboration),
  Phys.\ Rev.\  D {\bf 81}, 111101 (2010).
\bibitem{Belle:leta}
  Y.~Miyazaki {\it et al.}  (Belle Collaboration),
  Phys.\ Lett.\  B {\bf 648}, 341 (2007).
\bibitem{BaBar:leta}
  B.~Aubert {\it et al.}  (BaBar Collaboration),
  Phys.\ Rev.\ Lett.\  {\bf 98}, 061803 (2007).
\end{thebibliography}
\end{document}